\documentclass[12pt]{article}
\usepackage{amsmath}
\usepackage{graphicx}

\newcommand{\dd}{{\rm d}}
\begin{document}
\title{Interfacial Tension of  Electrolyte Solutions }
\author {Yan Levin \\
Instituto de F\'{\i}sica, Universidade Federal
do Rio Grande do Sul\\ Caixa Postal 15051, CEP 91501-970, 
Porto Alegre, RS, Brazil}

\maketitle
\begin{abstract}

A theory is presented to account for the increase in
surface tension of water in the presence
of electrolyte.  Unlike the original ``grand-canonical'' 
calculation of Onsager and Samaras,
which relied on the Gibbs adsorption isotherm and lead to a 
result which could only be expressed as an infinite series,
our approach is ``canonical''  and produces an analytic 
formula for the excess surface tension.
For small concentrations of electrolyte,  
our result reduces to the Onsager-Samaras limiting law.

\end{abstract}

\newpage

\section{Introduction}

It has been known for almost a century that aqueous solutions
of inorganic salts have greater surface tension than pure water \cite{hey10}.
An explanation of this effect was advanced by Wagner \cite{wag24} 
in $1924$ on the
the basis of the theory of strong electrolytes, which was introduced only
a year earlier by Debye and H\"uckel \cite{deb23}.  The fundamental insight
of Wagner was to realize that the presence of ions polarizes the
air-water interface, inducing an effective surface charge.  
Since the dielectric 
constant of water is significantly larger than that of air,
each ion's image charge equals to it in sign and magnitude.  
Thus, the repulsive 
interaction with the images {\it reduces} the density of electrolyte near
the interface. Appealing to the Gibbs adsorption isotherm \cite{gib28}, 
it is evident
that the depletion of solute near the interface results in an {\it increase}
of surface tension. 
Furthermore, the experimental observation that  for small
concentrations this increase  
 depends only  weakly on the ionic size \cite{hey10,lon42,ran63,ral73,wei96},
suggest the existence of a universal limiting law,
similar to the one obtained by Debye and H\"uckel for bulk 
properties of electrolyte solutions \cite{deb23}.  Indeed, the calculation
of Onsager and Samaras $(OS)$ seems to confirm such a limiting
law \cite{ons34}.  A number of approximations adopted
by $OS$  in the course of calculations, however, obscure
the full range of validity of their findings. To check the thermodynamic
consistency of the $OS$ results it is, therefore,
worthwhile to explore other routes to
surface tension.  In the absence of an exact calculation, 
these will provide a way to asses the
self-consistency of the formulas obtained.  
To this end, we propose a new approach for calculating 
the increase
in interfacial tension of water due to 1:1 electrolyte.
Our method differs from that of $OS$ --- who integrated the Gibbs adsorption
isotherm --- in that we  
identify the excess surface tension 
directly with the Helmholtz free energy necessary to create an interface.  
The advantage of this approach is that it allows to
write an analytic formula for the excess surface tension.  This
should be contrasted with the method of $OS$, who were able to express
their result only as an infinite sum. Nevertheless, in the
limit of large dilution our expression reduce to that of $OS$, 
suggesting that the limiting law  is,
indeed, exact. The paper is organized as follows:  in Sec.~\ref{thermo}
we briefly review the thermodynamics of interfaces and the relationship
between the ``canonical'' and the ``grand-canonical''  routes 
to surface tension;  in Sec.~\ref{tension} we outline the $OS$ theory
 and present our calculations;  
the results and conclusions are summarized in
Sec.~\ref{results}.

\section{Thermodynamics of Interfaces}
\label{thermo}

Consider an $r$ component mixture
confined to volume $V$.  
The system is in contact with a 
hypothetical reservoir of solute particles at temperature $T$ and
chemical potential $\mu_i$, with $\{i=1...r\}$. 
If the periodic boundary conditions are imposed
on $V$,   
at equilibrium, the 
system will have $\{N_i\}$ particles {\it uniformly} distributed 
throughout the volume.

On the other hand if $V$ forms part of a larger system,
its domain must be delimited by an interface. 
We shall idealize this
interface
as a mathematical surface --- the Gibbs dividing surface for solvent.
The discontinuity produced by the interface will affect
the interactions between the particles in its vicinity.
If the interface-particle interaction
is repulsive, it will lead to the depletion of solute
from the interfacial region, forcing it back {\it into} reservoir.
On the other hand, if this interaction is attractive, 
a concentrated layer of solute will build up 
along  the interface, producing
a net flow of solute {\it from} the reservoir. When the equilibrium is
reestablished, the distribution of solute is  no longer uniform,
but characterized by a density profile $\rho_i(z)$, where $z$
is a distance from the interface.  Clearly if the
system has a thermodynamic limit, the interface
does not influence the bulk distribution of particles and 
$\rho_i(\infty)=\rho_i$,
where $\rho_i=N_i/V$. The presence of an interacting interface, however, 
is responsible for
a net increase or decrease of solute in the system. 
Thus, we define the
amount of solute ``adsorbed''  as 
\begin{equation}
\label{1a}
N_i^s=\int_0^\infty \left(\rho_i(z)-\rho_i \right) \dd z.
\end{equation}
Note that this quantity can be either positive or negative depending
on whether the solute enters or leaves the system. Now, from
general thermodynamic principles,
the change in the total Helmholtz free energy of a system at 
{\it fixed volume, temperature, and the amount of solvent}
is, 
\begin{equation}
\label{1}
dF=\sigma dA + \sum_{i=1}^r \mu_i\, dN_i,
\end{equation}
where $\sigma$ is the surface tension and $A$ is the area of the interface.
We shall now divide the solute particles  into the
 ``bulk'' $N_i^b$ and the  ``surface'' $N_i^s$. 
Eq.~\ref{1} can  also be separated into the bulk,
\begin{equation}
\label{2}
dF^b=\sum_{i=1}^r \mu_i \, dN_i^b \;,
\end{equation}
and the surface contribution,
\begin{equation}
\label{3}
dF^s=\sigma dA + \sum_{i=1}^r \mu_i \, dN_i^s \;.
\end{equation}
Since $dF^s$ is an extensive function of $A$ and $\{N_i^s\}$, the
Euler's theorem for first-order homogeneous functions allows us to
integrate Eq.~\ref{3} yielding,
\begin{equation}
\label{4}
F^s=\sigma A + \sum_{i=1}^r \mu_i N_i^s.
\end{equation}
On the other hand, differentiating Eq.~\ref{4} and comparing it
with Eq.~\ref{3}, we find a Gibbs-Duhem-like equation,
\begin{equation}
\label{5}
A d\sigma + \sum_{i=1}^r N_i^s d\mu_i=0.
\end{equation}
For $r=1$ this reduces to the 
Gibbs adsorption isotherm,
\begin{equation}
\label{6}
\left. \frac{\partial\sigma}{\partial\mu} \right|_{T,V}=-\frac{N^s}{A}.
\end{equation}
Since the thermodynamic stability requires $d\mu/d\rho>0$, where
$\rho$ is the concentration of solute, it is evident that a positive
adsorption leads to a decrease
in surface tension, while a negative adsorption increases the surface
tension.  This, then, explains the Wagner's original observation
that the repulsion of ions from a polarized
air-water interface results in a depletion of electrolyte 
and an increase in surface tension.

Knowledge of ${N_i^s}$ allows a
calculation of the excess surface tension through the 
integration of Eq.~\ref{5}.  
Following Wagner, this was the procedure adopted by $OS$. 

The discussion outlined above relies on the presence of a hypothetical 
reservoir.  In
the language of statistical mechanics it is intrinsically ``grand-canonical''.
A different, ``canonical'' calculation should also be possible. In the
thermodynamic limit, the choice of ensemble will not matter,
if an {\it exact} calculation is performed.  
In practice, however, no exact calculation is
possible and approximations have to be made.  Thus, there is 
no {\it a priory}
guarantee that the two ensembles will lead to identical results. The
canonical approach presented bellow, is conceptually simpler
than its grand-canonical counterpart, since no reservoir is present.
Thus, when the interactions 
between the interface and the solute are turned on, 
the particles do not leave the systems and  $N_i^s=0$. 
Therefore, canonically the interfacial tension 
is equivalent to the
surface Helmholtz free energy density, $\sigma=F^s/A$. Now, consider
a mixture confined to a cylinder of length $H$ and a cross-sectional
area $A$.  Define $F$ as the total Helmholtz free energy of solute 
and $F^{bulk}$ as the free energy of solute in the absence
of an interface.  The change in the 
surface tension of solvent due to addition of solute is then
\begin{equation}
\label{7}
\sigma^{ex}=\lim_{A \rightarrow \infty}\frac{1}{A}
\lim_{H \rightarrow \infty}\left(F-F^{bulk}\right).
\end{equation}

\section{The Surface Tension}
\label{tension}

We are interested in the  surface tension of an interface between
an aqueous solution of a 
symmetric 1:1 electrolyte and air.  The extension to
asymmetric electrolytes is, in principle, straight forward.  
Some extra care, however, has to be taken to account for the strong
correlations between the cations and the anions, which result from
an increased ionic charge \cite{fis93,wei00}. In view of the
experimental observation that for small concentrations of
electrolyte the excess surface tension depends
only weakly on the ionic size \cite{hey10,ral73,wei96}, 
to simplify the calculations we
shall treat ions as point-like. The solvent will be modeled
as a uniform dielectric medium.

According to statistical mechanics, the concentration of solute
a distance $z$ from the interface is given by the 
Boltzmann distribution,
\begin{equation}
\label{8}
\rho_i(z)=\rho_i e^{-\beta W_i(z)} \;,
\end{equation}
where $\beta=1/k_BT$, and $W_i(z)$ is the adsorption potential
of the specie $i$ and is defined as the 
work required to bring a particle from infinity to distance $z$
from the interface.  For symmetric electrolyte $W_+(z)=W_-(z) \equiv W(z)$,
and $\rho_+(z)=\rho_-(z) \equiv \rho(z)$. 
Now, consider an ion located at distance $z$ from the interface.  If
the electrolyte is infinitely dilute, the electrostatic potential distance
$r_1$ from the ion can be calculated directly from the Laplace equation,
\begin{equation}
\label{9}
\psi(r_1,r_2)=\frac{q}{D r_1}+\frac{(D-D')}{(D+D')} \frac{q}{Dr_2} \;,
\end{equation}
where $D$ is the dielectric constant of water, $D'$ is the
dielectric constant of air,  and $r_2$ is the distance from the image charge
located opposite to the ion at $-z$. From the second term of Eq.~\ref{9}, 
the charge of the ``image ion'' is $q_{image}=(D-D') q/(D+D')$.
For an aqueous solution close to room temperature $D \approx 80$ and
$D'\approx 1$, so that
$q_{image} \approx q$. Therefore,  in a perturbative
theory, $D'/D$ can play a role of a small parameter.   
Since $D \gg D'$, the zeroth order calculation is already
quite accurate, and $D'$ can be set to zero. 
This is the first approximation proposed
by Wagner \cite{wag24} and  used  by $OS$ \cite{ons34}.  In the following
discussion we shall also adopt this approximation. The repulsive force
felt by an ion due to the dielectric discontinuity
produced by an air-water interface is then
\begin{equation}
\label{10}
{\cal F}(z)=\frac{q^2}{4Dz^2}.
\end{equation}
The amount of work required to bring this ion from infinity to a distance $z$ 
from the interface is 
\begin{equation}
\label{11}
W_{\infty}(z)=-\int_{\infty}^z {\cal F}(x)dx=\frac{q^2}{4Dz} \;,
\end{equation}
where we have added subscript $\infty$ to $W(z)$ to emphasize that the
calculation is done at infinite dilution. Alternatively, 
the G\"untelberg charging process \cite{gun26} can be used to calculate
the amount of electrostatic work necessary to ``create'' an ion of charge 
$q$ at distance $z$ from the interface,
\begin{equation}
\label{10a}
 W_\infty (z)= \int_0^1 \frac{\lambda q}{2 D z}\, q \dd \lambda =
\frac{q^2}{4 D z} \;.
\end{equation}

If the electrolyte is at finite concentration, the electrostatic
potential in the vicinity of a fixed ion satisfies the Debye-H\"uckel
equation,
\begin{equation}
\label{12}
\nabla^2 \psi=\kappa^2 \psi,
\end{equation}
where $\kappa^2(z)=8\pi\rho(z)/Dk_BT$. In order to simplify 
the calculations Onsager and Samaras suggested replacing 
$\kappa(z)$ in Eq.~\ref{12} by its bulk  value
$\kappa(\infty)$ \cite{ons34}. This, certainly, seems like a reasonable
thing to do in view of the fact that for an aqueous solution at
room temperature, the boundary layer is very narrow and
the density profile rapidly approaches its bulk value. 
Nevertheless, this approximation introduces some internal inconsistency 
into the theory which
can, in principle, manifest itself when different thermodynamic 
routes are taken to calculate the excess surface tension.  
To zeroth order in $D'/D$ and with $\kappa(z) \rightarrow \kappa(\infty)$, 
Eq.~\ref{12} can be 
integrated \cite{ons34,sti61} yielding
\begin{equation}
\label{14}
\psi(r_1,r_2)=\frac{q e^{-\kappa r_1}}{Dr_1}+\frac{q e^{-\kappa r_2}}{Dr_2},
\end{equation}
where, once again, $r_1$ is the distance from the ion, and $r_2$ is
the distance from the image charge located at $-z$.
The adsorption potential is obtained through the
G\"untelberg charging process or by direct force integration producing
\begin{equation}
\label{15}
W(z)=\frac{q^2 e^{-2 \kappa z}}{4 D z}.
\end{equation}
The density profile for cations and anions 
is found by substituting this expression
into Eq.~\ref{8},
\begin{equation}
\label{16}
\rho_\pm(z)=\rho \exp\left(\frac{be^{-2 \kappa z}}{2z}\right).
\end{equation}
We have defined $b$ as half the Bjerrum length, 
$b=\lambda_B/2=q^2/2Dk_BT$. For water at room temperature 
$b\approx 3.6$ \AA ---
comparable to the size of a hydrated ion --- and the density
profile rapidly reaches its bulk value. 
The amount of solute adsorbed
can now be calculated by inserting Eq.~\ref{16} into Eq.~\ref{1a}. With 
skillful complex variable analysis, $OS$ were able to integrate the
Gibbs adsorption isotherm, obtaining an expression for the excess 
surface tension
as an infinite series in $\kappa b$ \cite{ons34}.
As was discussed in Sec.~\ref{thermo}, this procedure does not 
conserve the 
number of particles in the system and is,
intrinsically, grand-canonical.   
We now present an alternative, canonical, calculation of the 
excess surface tension.  

Lets suppose that the electrolyte is confined to a cylinder of height
$H$ and cross-sectional area $A$.  If the interface-ion interactions are
neglected (periodic boundary conditions), 
the electrolyte will be  uniformly distributed over the volume
of the cylinder.  In the thermodynamic limit, the 
electrostatic free energy can be easily calculated
from the Debye-H\"uckel theory yielding,
\begin{equation}
\label{17}
F^{bulk}=-\frac{q^2 \kappa}{3 D}N_t,
\end{equation}
where $N_t=N_++N_-$ is the total number of solute particles.  
On the other hand, the presence of an interface 
produces a concentration gradient characterized by the 
{\it normalized} distribution,
\begin{equation}
\label{18}
\rho_\pm(z)=\frac{N_\pm e^{-\beta W(z)}}
{A \int_0^H e^{-\beta W(z)} \dd z}.
\end{equation}
Now, suppose we fix an ion some distance $z$ from the interface,
far from the radial boundary of the cylinder. 
The electrostatic potential in the vicinity of this ion is approximately
given by Eq.~\ref{14} --- where in view of the thermodynamic limit of
Eq.~\ref{7}, we can neglect
the finite size corrections.   Evidently this potential is produced by
the ion itself, as well as, by the interface and the ionic atmosphere.
The potential that the ion feels due to polarization of the
interface and the ionic atmosphere is,
\begin{equation}
\label{19}
\psi_0(z)=\lim_{\substack{r_1 \rightarrow 0 \\ r_2 \rightarrow 2z}} 
\left(\psi(r_1,r_2)-\frac{q}{Dr_1}\right)=\frac{-q\kappa}{D}+
\frac{qe^{-2\kappa z}}{2Dz}.
\end{equation}
The electrostatic energy is,
\begin{equation}
\label{20}
E=\frac{Aq}{2}\int_0^H \left(\rho_+(z)+\rho_-(z) \right)\psi_0(z)dz.
\end{equation}
This can be subdivided into the bulk and surface contributions, 
corresponding to the first and the second 
term of Eq.~\ref{19}, respectively.
The bulk term is easily integrated yielding,
\begin{equation}
\label{21}
E^b=\frac{-q^2\kappa}{2D}N_t.
\end{equation}
The surface contribution is found to be,
\begin{equation}
\label{22}
E^s=\frac{q^2 N_t}{4D}\frac{\displaystyle \int_0^H 
\exp\left(-2\kappa z-\beta W(z)\right)\frac{dz}{z}}
{\displaystyle \int_0^H \exp\left(-\beta W(z)\right) dz}\;.
\end{equation}
Now, in the limit $H \rightarrow \infty$ relevant for the calculation of
surface tension,
\begin{equation}
\label{23}
\lim_{H \rightarrow \infty} 
\frac{1}{H}\int_0^H \left(e^{-\beta W(z)}-1\right) dz =0,
\end{equation}
and the Eq.~\ref{22} simplifies to,
\begin{equation}
\label{24}
E^s=\frac{A q^2 \rho}{2D}\int_0^\infty e^{-2\kappa z-\beta W(z)}\frac{dz}{z} \;.
\end{equation}
Unfortunately, no method is available for analytic 
evaluation of this integral. 
We note, however, that if the adsorption potential
is replaced by its value at infinite dilution, $W(z) \rightarrow W_\infty(z)$,
the integral can be done explicitly yielding,
\begin{equation}
\label{25}
E^s=\frac{q^2 \rho A}{D}K_0\left(2 \sqrt{b\kappa}\right),
\end{equation}
where $K_\nu(x)$ is the modified Bessel function of order $\nu$. Replacement
of the adsorption potential by its value at infinite dilution should
be a good approximation, since the large distances --- 
for which the discrepancy between
$W(z)$ and $W_\infty(z)$ becomes significant --- are not important because
of the exponential drop in the electrostatic potential away from the
interface. To confirm this, we have numerically evaluated
the integral
\begin{equation}
\label{25a}
I(\kappa b)=\int_0^\infty e^{-2\kappa z-\beta W(z)}\frac{dz}{z} \;
\end{equation}
and compared it with the exact analytic expression obtained when
$W(z) \rightarrow W_\infty(z)$, Fig. 1.  As was hoped
the agreement is, indeed, quite good extending all
the way to $y\equiv \kappa b\approx .45$. In water
at room temperature, this corresponds to
concentration of $0.15 M$, which is 
above the maximum for which the limiting
laws of this paper can be realistically expected to apply.   
\begin{figure}[h]
\includegraphics[width=9cm,angle=270]{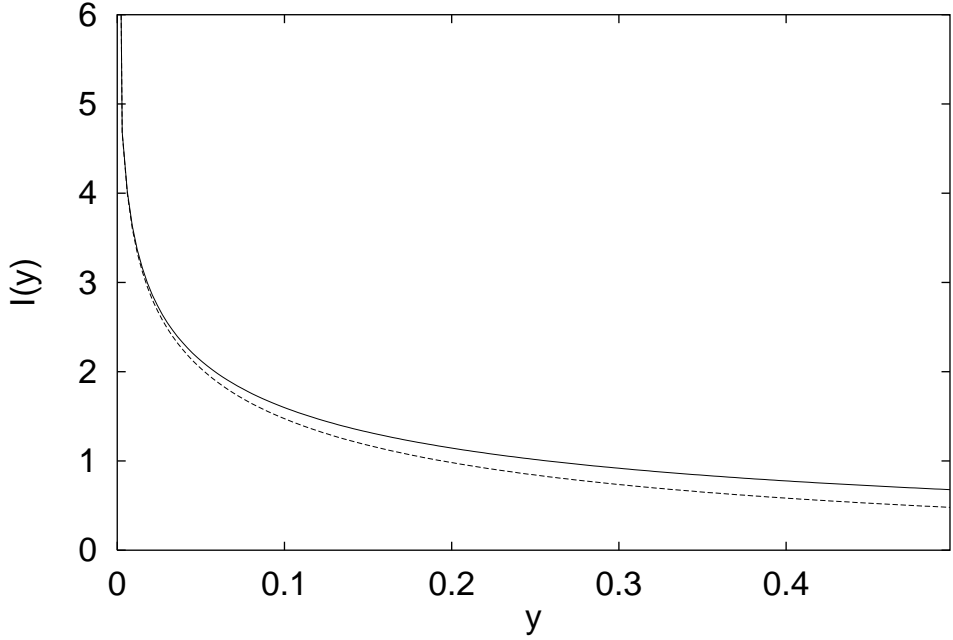}
\caption{The solid curve is the numerically evaluated integral $I(y)$, 
as a function of $y=\kappa b$, Eq.~\ref{25a}; the dashed curve is the 
analytic expression obtained when $W(z) \rightarrow W_\infty(z)$, see
Eq.~\ref{25}}
\label{Fig1}
\end{figure}

The electrostatic {\it free energy} can now be obtained
through the Debye charging process, 
in which {\it all} the particles are simultaneously
charged from zero to their full  charge\cite{deb23,fis93},
\begin{equation}
\label{26}
F=\int_0^1 2E(\lambda q)\frac{\dd \lambda}{\lambda},
\end{equation}
note that both $b$ and $\kappa$ are dependent on $q$, so
that $b(\lambda q)=\lambda^2 b(q)$ and $\kappa(\lambda q)=\lambda \kappa(q)$.
The integral can, once again, be done explicitly yielding the electrostatic
free energy of an electrolyte solution in the presence of an
interface,
\begin{eqnarray}
\label{27}
F&=&-\frac{q^2\kappa}{3D}N_t-\frac{Aq^2\rho}{2D} \left[2K_0(2\sqrt{y})
\;{}_1F_2(1;2/3,5/3;y)+ \right. \nonumber \\
&& \left.3\sqrt{y} K_1(2\sqrt{y})\;{}_1F_2(1;5/3,5/3;y)\right].
\end{eqnarray}
Here, ${}_pF_q$ is the generalized hypergeometric function \cite{bai35} and
$y \equiv b\kappa$. Substituting Eq.~\ref{27}
into Eq.~\ref{7}, the increase in surface tension of water due to  
 1:1 electrolyte is,
\begin{eqnarray}
\label{28}
\sigma^{ex}=\sigma_0^{ex}\left[2K_0(2\sqrt{y})
\;{}_1F_2(1;2/3,5/3;y)+
3\sqrt{y} K_1(2\sqrt{y})\;{}_1F_2(1;5/3,5/3;y)\right],
\end{eqnarray}
where $\sigma_0^{ex}=q^2\rho/2D$.
For very low concentrations, this expression reduces to the limiting law
found by $OS$,
\begin{eqnarray}
\label{29}
\sigma^{ex}_l=\sigma_0^{ex}[-\ln(y)-2\gamma_E+3/2],
\end{eqnarray}
where $\gamma_E=0.57721566490153...$ is the Euler's constant.

\section{Results and Discussion} 
\label{results}

We have presented a ``canonical'' calculation of the
excess surface tension in an electrolyte solution. Unlike
the earlier ``grand-canonical'' 
method of  Onsager and Samaras, 
our approach leads to an analytic expression for the excess surface
tension expressed in terms of Bessel and hypergeometric functions. 
It is gratifying, however, that in spite of all the approximations, 
the two ensembles 
produce  the identical limiting law.  This  
thermodynamic self-consistency 
suggests that the $OS$ limiting law is, indeed, exact to zeroth order in
$D'/D$. 
In Fig. 2 we plot, as a function of
concentration,  the $\sigma^{ex}$ Eq.~\ref{28}; the sum of the
first twenty terms of the infinite series for the excess surface tension
obtained by $OS$ \cite{ons34};  and the $OS$ limiting law Eq.~\ref{29}.
The surface tension $\sigma$ is measured
in $mN \cdot m^{-1}$  and the concentration of salt $c$ in 
moles/liter $(M)$ so that,
\begin{eqnarray}
\label{30}
\sigma^{ex}_0&=&\frac{69.4692 c}{D}\left(mN \cdot m^{-1} \right) \, \nonumber \\
y&=&\frac{4201742\sqrt{c}}{(DT)^{3/2}}.
\end{eqnarray}
For water at room temperature $D\approx 78.54$. 
\begin{figure}[h]
\includegraphics[width=9cm,angle=270]{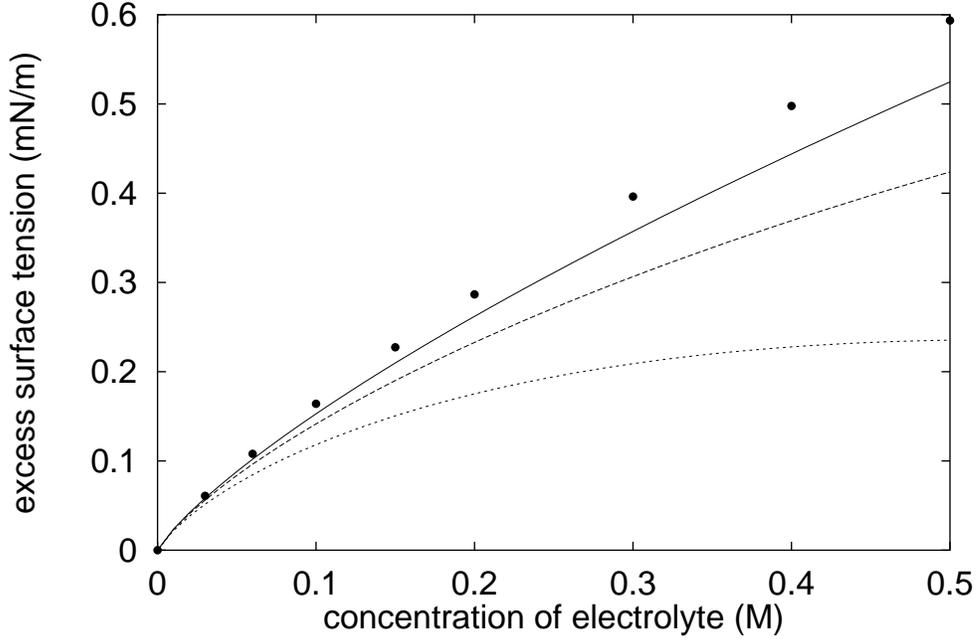}
\caption{The excess surface tension of an aqueous solution of
a 1:1 electrolyte at room temperature.  The solid curve is the analytic
expression given by Eq.~\ref{28}; the long dashed curve is the sum of
the first 20 terms of the $OS$ result \cite{ons34}; the short dashed curve 
is the Onsager-Samaras limiting law, Eq.~\ref{29}; the solid points
are the result of numerical integration of Eq.~\ref{26} with full 
$W(z)$.}
\label{Fig2}
\end{figure}
We note that both
ensembles agree fairly well over the full range of concentrations,
with the ``canonical'' calculation predicting a somewhat
larger excess surface tension.  
In Fig. 1 it was shown  that the substitution 
$W(z) \rightarrow W_\infty(z)$ leads to a  good approximation
for the electrostatic energy.  To confirm that
this also extends to the {\it free} energy,
we have performed the Debye charging process, Eq.~\ref{26}, numerically
using the explicit form of $W(z)$, Eq.~\ref{15}. The result
is plotted as solid points in Fig. 2. 
Indeed, the numerically calculated surface
tension agrees quite well with the analytic result, Eq.~\ref{28},
obtained with the substitution $W(z) \rightarrow W_\infty(z)$.

It has been noted
that the $OS$ theory gives a fairly good quantitative description for
concentrations up to $0.1M$, above which it consistently underestimates
the increase in the interfacial tension \cite{lon42,ran63,ral73,wei96}. 
The canonical calculation presented above extends the range
of agreement between theory and experiment.  
It is, however, unrealistic to demand that the theory presented
above should apply to concentrated solutions, for which
even the bulk thermodynamic 
properties loose their universality.  Thus, for concentrations
above $0.2M$, the molecular nature of the solvent as well as the 
lyotropic properties of solute will become important.
In fact, it has been observed experimentally that for 
concentrated solutions the excess surface tension increases {\it linearly}
with the concentration of electrolyte. For an aqueous 
solution of  $NaCl$ at $T=25^oC$ it is found that 
$\sigma^{ex} \approx 1.6 c$ \cite{mat98}, which begins to show
strong deviation from Eq.~\ref{28} for $c>0.2M$.

\section{acknowledgments}
This work was supported in part by CNPq --- 
Conselho Nacional de
Desenvolvimento Cient{\'\i}fico e Tecnol{\'o}gico and FINEP --- Financiadora 
de Estudos e Projetos,
Brazil.

\newpage
\section{Figure Captions}

Fig. 1: The solid curve is the numerically evaluated integral $I(y)$, 
as a function of $y=\kappa b$, Eq.~\ref{25a}; the dashed curve is the 
analytic expression obtained when $W(z) \rightarrow W_\infty(z)$, see
Eq.~\ref{25}

Fig. 2: The excess surface tension of an aqueous solution of
a 1:1 electrolyte at room temperature.  The solid curve is the analytic
expression given by Eq.~\ref{28}; the long dashed curve is the sum of
the first 20 terms of the $OS$ result \cite{ons34}; the short dashed curve 
is the Onsager-Samaras limiting law, Eq.~\ref{29}; the solid points
are the result of numerical integration of Eq.~\ref{26} with full 
$W(z)$.

\end{document}